\RequirePackage{lineno}
\documentclass[aps,prd,twocolumn,superscriptaddress,amsmath,amssymb]{revtex4}

\usepackage{graphicx} 
\usepackage{epstopdf} 
\usepackage{dcolumn} 
\usepackage{xcolor} 
\usepackage{amsmath,amssymb} 
\usepackage{hyperref} 
\usepackage[utf8]{inputenc} 
\usepackage[english]{babel} 
\usepackage{blindtext} 
\usepackage{ifthen} 
\usepackage{orcidlink} 

\usepackage{mwe,tikz}
\usepackage[percent]{overpic}

\usepackage[italic]{hepnames}
\usepackage[separate-uncertainty=true]{siunitx}

\graphicspath{{figures/}} 

\newboolean{articletitles}
\setboolean{articletitles}{true} 

\input{belle2-symbols}

\begin{document}

\keywords{CKM, B meson, CP violation, Time depedent, qqs decay, etaprime}
\preprint{Belle II Preprint 2024-003}
\preprint{KEK Preprint 2023-50}

\title{Measurement of \CP asymmetries in $\PBz\to\Petaprime\PKs$ decays at Belle II
}
  \author{I.~Adachi\,\orcidlink{0000-0003-2287-0173}} 
  \author{L.~Aggarwal\,\orcidlink{0000-0002-0909-7537}} 
  \author{H.~Ahmed\,\orcidlink{0000-0003-3976-7498}} 
  \author{H.~Aihara\,\orcidlink{0000-0002-1907-5964}} 
  \author{N.~Akopov\,\orcidlink{0000-0002-4425-2096}} 
  \author{A.~Aloisio\,\orcidlink{0000-0002-3883-6693}} 
  \author{N.~Anh~Ky\,\orcidlink{0000-0003-0471-197X}} 
  \author{D.~M.~Asner\,\orcidlink{0000-0002-1586-5790}} 
  \author{H.~Atmacan\,\orcidlink{0000-0003-2435-501X}} 
  \author{T.~Aushev\,\orcidlink{0000-0002-6347-7055}} 
% ...
\collaboration{The Belle II Collaboration}

\date{February 5, 2024}
\begin{abstract}

We describe
a measurement of charge-parity (\CP) violation asymmetries in
$\PBz\to\Petaprime\PKs$ decays using Belle II data. 
We consider $\Petaprime\to\Peta(\to\Pphoton\Pphoton)\Pgpp\Pgpm$ and
$\Petaprime\to\Prho(\to\Pgpp\Pgpm)\Pphoton$ decays.
The data were collected at the SuperKEKB asymmetric-energy $e^+e^-$ collider between the years 2019 and 2022,
and contain $(387\pm 6) \times 10^6$ bottom-antibottom meson pairs.
We reconstruct $829\pm35$ signal decays
and extract the \CP violating parameters from a fit to the distribution 
of the proper-decay-time difference between the two \PB mesons.
The measured direct and mixing-induced 
\CP asymmetries are $\CetapKs = -0.19 \pm 0.08 \pm 0.03 $ and $\SetapKs = +0.67 \pm 0.10 \pm 0.03 $,
respectively, where the first uncertainties are statistical and the second are systematic.
These results are in agreement with current world averages and standard model predictions.

\end{abstract}

\maketitle

\section{Introduction}
In the standard model (SM), the only source of charge-parity (\CP) violation is an irreducible phase in the Cabibbo-Kobayashi-Maskawa (CKM) quark-mixing matrix~\cite{PhysRevLett.10.531, Kobayashi:1973fv}.
This phase is measured with high precision in tree-dominated 
$b\to{c\bar{c}s}$ decays~\cite{PhysRevLett.108.171802, PhysRevD.79.072009, PhysRevLett.115.031601}, \eg, 
$\PBz\to\PJpsi\PKz$.
In contrast, $b\to{s}q\bar{q}$ decays, with $q$ indicating $u$, $d$, or $s$ quark, are dominated by loop amplitudes, in which additional sources of \CP violation from physics beyond the SM could be involved~\cite{Kou:2018nap, Bevan:2014iga}. 
A comparison between \CP asymmetries measured precisely in $b\to{s}q\bar{q}$ and $b\to{c\bar{c}s}$ transitions can thus probe non-SM physics.

The $\PBz\to \Petaprime \PKs$ decay is of particular interest
due to its relatively large branching fraction 
and limited contribution from tree amplitudes compared to other  $b\to{s}q\bar{q}$ decays.
The deviation of the mixing-induced \CP asymmetry (\SetapKs) from $\sin2\phi_1$ is expected to be $0.01\pm0.01$~\cite{Beneke:2005pu}, in the SM, where $\phi_1\equiv\arg{(-V_{cd}V^*_{cb}/V_{td}V^*_{tb})}$ is an angle of the CKM unitarity triangle
and $V_{ij}$ are the CKM matrix elements.
The direct \CP asymmetry (\CetapKs) is predicted to be zero~\cite{pdg_CKM}.

Measurements of \CetapKs and \SetapKs have been reported by the Belle~\cite{Belle:2014atq} and 
BaBar~\cite{BaBar:2008ucf} experiments, yielding the current world averages
$\CetapKs = -0.05 \pm 0.04$ and $\SetapKs = 0.63 \pm 0.06$~\cite{HFLAV:2022pwe}.
These are the most precise measurements of \CP asymmetries  with
 $b\to sq\bar{q}$ transitions.
However, improved measurements are needed to match the precision of 
the theoretical prediction of possible deviations from the SM~\cite{Beneke:2005pu}.
The sensitivity of experiments operating at hadron colliders, such as LHCb, is 
limited by the challenge of the reconstruction of neutral final-state particles.

At an $e^+e^-$ flavor-factory, \BBbarz pairs are produced via the process $e^+e^- \to \PupsilonFourS\to\BBbarz$.
We consider the case when one neutral \PB meson (\Btag) decays into a flavor-specific final state at time $t_\text{tag}$, and 
the other \PB (\Bcp) decays into a \CP eigenstate at time $t_\textsl{CP}$.
As the two neutral \PB mesons remain in a quantum-entangled state until one of them decays, 
the flavor of \Bcp is opposite to that of \Btag at $t_\text{tag}$.
We define $\q$ as the flavor of \Btag at $t_\text{tag}$, with $\q$ taking the value $+1 \; (-1)$ for \PBz  (\PaBz ).
The decay rate for the \Bcp can be given by

\begin{equation}\label{eq:P}
  \mathcal{P}(\Delta{t},\q) = \frac{e^{-|\Delta{t}|/\tau_{B^0}}}{4\tau_{B^0}}\left[1+\q\mathcal{A}_{\CP}(\Delta{t})\right],
\end{equation}
where $\Delta{t}=t_\textsl{CP}-t_\text{tag}$ is the proper-time difference between the \Bcp and \Btag decays,
$\tau_{B^0}$ is the \PBz lifetime, and $\mathcal{A}_{\CP}$ is the time-dependent \CP asymmetry, defined as

\begin{equation}\label{eq:asym}
    \begin{split}
    \mathcal{A}&_{\CP}(\Delta{t}) = \frac{\Gamma(\PaBz\to\Petaprime\PKs)-\Gamma(\PBz\to\Petaprime\PKs)}{\Gamma(\PaBz\to\Petaprime\PKs)+\Gamma(\PBz\to\Petaprime\PKs)} \\
    &= \SetapKs\sin(\Delta{m_d}\Delta{t})-\CetapKs\cos(\Delta{m_d}\Delta{t}),
\end{split}
\end{equation}
where  $\Delta{m_d}$ is the mass difference between the two neutral \B-meson mass eigenstates.

This paper reports a measurement of \CP asymmetries
\CetapKs and \SetapKs, based on the data collected by the Belle II experiment in 2019--2022 at
the SuperKEKB asymmetric-energy $e^+e^-$ collider~\cite{Akai:2018mbz},
operating at the  \PupsilonFourS resonance.
The total integrated luminosity is \num{362\pm2}\invfb, which corresponds to \num{387\pm6 e6} \BBbar pairs.

We reconstruct the signal decay (\Bcp) by combining the $\PKs\to\Pgpp\Pgpm$ candidate with the \Petaprime meson reconstructed in two channels, 
$\Petaprime\to\Peta(\to\gamma\gamma)\Pgpp\Pgpm$ and $\Petaprime\to\Prho(\to\Pgpp\Pgpm)\Pphoton$.
We also explore the channel $\Petaprime\to\Peta(\to\Pgpm\Pgpp\Pgpz)\Pgpp\Pgpm$, but exclude it from the final result due to its large statistical uncertainty.
The \Btag flavor is determined with a flavor tagging algorithm~\cite{Abudin_n_2022}.
The values of \CetapKs and \SetapKs are extracted via a maximum-likelihood fit to the distributions of $\Delta{t}$ and other observables that discriminate signal from background.
The analysis technique and the $\Delta{t}$ resolution model are tested on the $\PBp\to\Petaprime\PKp$ control channel, where we do not expect any \CP violation.
The measurement of the lifetimes of \PBz and \PBp mesons validates $\Delta{t}$ resolution modeling.
Charge-conjugated modes are implied unless otherwise specified.

\section{Belle II detector and simulation}
Belle II~\cite{Abe:2010gxa} is a particle physics experiment operating at the SuperKEKB collider
in Tsukuba, Japan.
Several subsystems, cylindrically arranged around the interaction point, enable reconstruction of heavy flavor particles
and \Ptau leptons produced in energy-asymmetric $e^+e^-$ collisions.
The innermost part of the detector comprises a two-layer silicon pixel
detector (PXD), surrounded by a four-layer double-sided silicon microstrip
detector (SVD). Together, they provide information about the charged particle trajectories (tracks)
and \PB decay positions (vertices). The momenta and charge of charged particles are reconstructed
with a 56-layer central drift chamber (CDC), which is the main tracking subsystem. 
Only one sixth of the second PXD layer
is installed for the data analyzed in this paper.

Charged particle identification (PID)
is accomplished by a time-of-propagation counter  and an aerogel ring-imaging Cherenkov
counter, located in the barrel and forward-endcap regions, respectively.
The CDC provides additional PID information through the measurement of specific ionization.
An electromagnetic calorimeter (ECL), made of CsI(Tl) crystals, is used for precise
determination of the photon energy and angular coordinates as well as for electron
identification.
The tracking, PID, and ECL subsystems are surrounded by a superconducting solenoid,
providing an axial magnetic field of $1.5\,\text{T}$.
A \KL and muon identification system is located outside of the magnet and consists of flux-return iron plates interspersed
with resistive plate chambers and plastic scintillators.
The central axis of the solenoid defines the $z$ axis of the laboratory frame, pointing approximately in the direction of the electron beam,
with respect to which the polar angle $\theta$ is defined.

The analysis strategy is tested and optimized on Monte Carlo simulated event samples before being applied to the data.
Simulation is also used to determine the signal efficiency and the fit model.
Quark-antiquark pairs from $e^+e^-$ collisions are generated using
\textsc{KKMC}~\cite{Jadach:1999vf} with \textsc{Pythia8}~\cite{Sjostrand:2014zea}, while hadron
decays are simulated with \textsc{EvtGen}~\cite{Lange:2001uf}. The detector response and
\PKs decays are simulated using \textsc{Geant4}~\cite{Agostinelli:2002hh}.
Both data and simulated samples are processed using the Belle II analysis software framework~\cite{Kuhr:2018lps, the_belle_ii_collaboration_2022_6949466}.

\section{Event reconstruction and selection}
The reconstruction of signal candidates starts by reconstructing the \PBz decay products,
$\Petaprime[\to\Peta(\to{\Pphoton\Pphoton})\Pgpp\Pgpm]$, $\Petaprime[\to\Prho(\to\Pgpp\Pgpm)\Pphoton]$, and $\PKs\to\Pgpp\Pgpm$.

Charged particles, assumed to be pions, are reconstructed using the tracking algorithm described in Ref.~\cite{Bertacchi:2020eez},
with measurement points from tracking subdetectors (PXD, SVD, and CDC).
Pions are required to be within the CDC angular acceptance ($17^{\circ} < \theta < 150^{\circ}$)
and to have a distance of closest approach from the interaction
point less than 2.0\,cm along the $z$ axis and less than 0.5\,cm in the transverse
plane, in order to reduce contamination from tracks not produced in the collision.
Furthermore, at least one of the two charged pions from \Petaprime  or \Prho,
which are used to reconstruct the \Bcp decay vertex, is required to have at least one PXD measurement point.

The photons are reconstructed from ECL energy deposits not associated to any track.
They are required to be in the CDC angular acceptance and have energy deposit in more than one ECL crystal.

The \PKs candidates are reconstructed from two oppositely charged pions
coming from the same vertex and required to
have a momentum direction compatible with the direction defined by the \PKs and \Bcp decay
vertices ($\cos\alpha>0.99$, where $\alpha$ is the angle between the two directions) and have an invariant mass $0.49<m(\Pgpp\Pgpm)<0.51\gevcc$.

For the first decay subchannel,  the $\Peta\to\Pphoton\Pphoton$ 
candidates are reconstructed from two photons with energies greater than $150\,\mev$ and an invariant mass in the range
$0.505<m(\gamma\gamma)<0.580\gevcc$. The candidate \Peta and two oppositely charged
pions are combined to form an \Petaprime, which is retained if its mass satisfies $0.945<m(\Peta\Pgpp\Pgpm)<0.970\gevcc$.
The \Peta mass is constrained to its known value~\cite{ParticleDataGroup:2022pth} to reconstruct the \Petaprime candidate.

For $\Petaprime\to\Prho\Pphoton$ subchannel, we require two charged pions to first form a \Prho candidate.
For this subchannel, the pions are required to satisfy a PID requirement,
computed using all PID capable detectors, which has a \Pgppm identification efficiency of about 90\%,  
and a \PKpm  mis-identification probability of about 10\%.
The pion tracks are also required to have at least 20 measurement points
in the CDC, which is sufficient to provide high efficiency and a low rate of  misreconstructed tracks.
The tighter selection for the pions in the second subchannel helps to reduce the larger background as well as the number of misreconstructed signal candidates.
A less restrictive requirement is applied on the dipion mass, due to the broad width of the \Prho resonance:
$0.51<m(\Pgpp\Pgpm)<1.0\gevcc$.
The lower bound of the invariant mass criterion avoids contamination from \KS\ decays.
An \Petaprime candidate is formed combining a \Prho and a photon candidate with energy greater than 250\mev .
We require the invariant mass to be in the range $0.92<m(\Prho\Pphoton)<0.98\gevcc$. 
This subchannel has broader \Petaprime mass resolution due to lack of constraint of \Prho mass.

The invariant mass criteria for the \Peta, \Petaprime, and \KS\ candidates correspond to approximately $\pm 1.7 \sigma_m$ intervals, where $\sigma_m$ is the Gaussian mass resolution of each decay mode.

The \Petaprime and \PKs candidates are combined to form a \PBz candidate. 
The \PBz vertex is determined by
fitting the entire decay chain with the {\tt TreeFitter} algorithm~\cite{Krohn:2019dlq, Hulsbergen_2005},
constraining the mass of all intermediate mesons to their known values \cite{ParticleDataGroup:2022pth}, except for
the \Prho, due to its large width, and requiring the fit to converge.
The momenta of all particles are updated after this vertex fit.
The \PBz vertex is constrained to point back, 
along the direction of the reconstructed \PBz momentum,
to the interaction region,
calibrated with $e^+e^-\to\mu^+\mu^-$.

For each \PB candidate, the beam-energy constrained mass 
$\Mbc=\sqrt{E^{*2}_\text{beam}/c^4-p^{*2}_B/c^2}$ and energy difference $\Delta{E}=E^{*}_B-E^{*}_\text{beam}$ are calculated,
where $(E,p)^{*}_B$ is the four-momentum of the \PB candidate and $E^{*}_\text{beam}$ is the beam energy,
both calculated in the center-of-mass frame.
For correctly reconstructed signal events, \Mbc peaks at the \PBz invariant mass and \De at zero. 
The requirements $\Mbc>5.2~\gevcc$ and $|\Delta{E}|<0.2~\gev$ are applied.

In data, the average \PB candidate multiplicity for events with at least one reconstructed candidate
is about 1.4 for \BzchInoSpace\ and 1.8 for \BzchTnoSpace. 
The difference is due to the presence of an intermediate, narrow resonance (\Peta) in the first subchannel.
If multiple $B$ candidates are present in an event, the one with the smallest \PBz vertex $\chi^2$ value is
retained. In the simulation, this criterion selects the correct candidate in more than $99\%$ of the cases when a true one is reconstructed.

The reconstruction efficiencies, determined using simulation, are $28.3\%$ and $19.2\%$ for \BzchInoSpace\ and \BzchTnoSpace, respectively.
The lower efficiency for the second subchannel is mostly due to PID requirements applied to the two pions from the \Prho decay in order to suppress background.

The control channel $\PBp\to\Petaprime\PKp$ uses an \Petaprime reconstructed as described above and a charged kaon candidate
with $\theta<136^\circ$, 
excluding candidates
in the backward part of the detector, where the background is higher. The PID selection for the \PKp candidate  
has an efficiency of about $90\%$ for kaons, and a misidentification probability for pions of about $5\%$.
In the control channel, the \PKp is not used for the \PBp vertex determination in order to have a vertex resolution similar
to that of the signal channel.

The \Btag candidate is reconstructed using all the charged particles that are not associated to \Bcp, having
measurement points both in the SVD and CDC, and a momentum greater than 50\mevc.
The RAVE algorithm is used to reconstruct the \Btag vertex~\cite{RAVE}.
This algorithm downweights tracks with large contributions to the vertex $\chi^2$, which are likely to originate from decays of secondary long-lived charm hadrons.
The decay position of \Btag is
determined by constraining its direction, as determined from its decay vertex and the interaction point, to be collinear
with its momentum vector, reconstructed from the \Bcp and momenta of the colliding beams~\cite{btube}.

The proper-decay-time difference $\Delta{t}$ between the two \PB mesons is determined from the positions of their
reconstructed vertices along the Lorentz  boost axis,
$\Delta{t}=\Delta{z}/\beta\gamma\gamma^*{c}$, where $\beta\gamma=0.287$ is the boost of the \PupsilonFourS
with respect to the laboratory frame, 
and $\gamma^*=1.002$ is the Lorentz factor of the \B meson in the center-of-mass frame.
We reject poorly reconstructed events requiring $|\Delta{t}|<8~\textrm{ps}$ and
require the per-event uncertainty $\sigma_{\Delta{t}}$ to be less than $2~\textrm{ps}$.

The main source of background is random combinations of tracks and photons that arise from continuum $e^+e^-\to q\bar{q}$ ($q=u,d,s,c$) events.
A boosted-decision-tree (BDT) classifier~\cite{Keck:2017gsv} is trained using 26 
event-shape variables to separate jet-like continuum events from more spherical \BBbar topologies.
The variables, in order of decreasing discriminating power, are the cosine of the angle
between the \Bcp and \Btag thrust axes~\cite{BRANDT196457}, the cosine of the angle between the \Bcp thrust axis and the $z$ axis, 
the ratio of the second to the zeroth Fox–Wolfram moments~\cite{PhysRevLett.41.1581}, 
the modified Fox–Wolfram moments~\cite{PhysRevLett.91.261801}, 
the \Btag thrust magnitude, and the CLEO cones~\cite{PhysRevD.53.1039}.
Variables that exhibit correlations greater than 10\% with those used for the \CP asymmetry measurement (\Mbc, \De, \Dt, and \sDt) are excluded.
The BDT training is performed using simulated signal  events and data in the sidebands ($\Mbc<5.27\gevcc$ 
and $\De<-0.07$ or $>0.05\gev$), which are dominated by continuum background.
As a consistency check, we repeat the training with the off-resonance data collected 60\mev below the 
\PupsilonFourS resonance, for an integrated luminosity of $42~\invfb$, and find the results to be in agreement with those obtained with the nominal training.
A high-efficiency selection on the BDT output \CS is applied, retaining about 95\% of the signal while suppressing 60\% of the continuum background.

\section{Time-dependent \CP asymmetry fit}
The parameters \CetapKs and \SetapKs are extracted with an extended unbinned 
maximum-likelihood fit using the \Mbc, \De, \CS, \Dt, and tag-flavor $\q$ observables,
plus $\sigma_{\Delta{t}}$ as a conditional observable for \Dt resolution function.
The first three observables provide discrimination between signal and continuum background.
The \Dt and $\q$ ones provide access to time-dependent \CP asymmetry.
Four different sample components are considered: signal;
self-cross-feed (SxF), where a signal decay is 
misreconstructed, mostly due to wrong \Pphoton associations; background from 
continuum; and background from \BBbar.
The SxF sample amounts to about 5\% of the signal while \BBbar amounts to about 1\% of the continuum.

The fit is performed in two steps.
In the first step the shapes and yields of all sample components are reliably determined in a time-independent fit,
using \Mbc, \De, and \CS distributions  in the region $\Mbc>5.2\gevcc$ and $|\De|<0.2\gev$,
which in turn simplifies the time-dependent fit of the second step.
In the first step most parameters of the signal and all those of the continuum models are allowed to vary, as well as the yields of signal and continuum.
The SxF shape is fixed from simulated events, as well as its normalization relative to the signal component.
The shape and yields of the \BBbar component are fixed from simulation, as their contribution is too small to be determined from data.

The second step of the fit uses \Dt and $\q$ in addition to the three observables utilized in the first step,
and is performed only in the signal region, defined by $\Mbc>5.27\gevcc$ and $-0.07<\De<0.05\gev$, where approximately $98\%$ of signal is present.
In this region the SxF amounts to about 3\% of signal.
In this step, the shapes for \Mbc, \De, and \CS as well as the yields of all components are fixed from the previous one,
so the only free parameters are \CetapKs and \SetapKs.

The \Btag flavor is determined using the flavor tagging algorithm described in Ref.~\cite{Abudin_n_2022},
which uses the properties of particles not associated with the \Bcp.
The algorithm provides the flavor $\q$ and the tagging quality $r=(1-2w)$, where $w$ is the mistagging probability.
The range of $r$ varies from $r=0$ (no flavor information can be obtained) to $r=1$ (corresponding to unambiguous flavour determination).

Both fit steps are performed simultaneously in seven subsets of data (bins) selected according to tagging quality $r$,
with boundaries set at 0, 0.1, 0.25, 0.45, 0.6, 0.725, 0.875, and 1, to gain statistical sensitivity from events with different wrong-tag fractions.
In the first step, the signal and continuum yields of each bin are varied independently.

Each observable is modeled independently and hence the total probability density function (PDF) 
is the product of the four independent PDFs,
as shown in Eq.~\ref{eq:totalPdf} and~\ref{eq:TIPDF}:
\begin{equation}\label{eq:totalPdf}
\begin{split}
\text{Model}(\Mbc,\De, & \CS, \Dt) =\\
&\mathcal{F}(\Mbc,\De,\CS)\cdot\mathcal{P}(\Dt),\\
\end{split}
\end{equation}
with:
\begin{equation}\label{eq:TIPDF}
\begin{split}
\mathcal{F}(\Mbc,& \De,\CS) =\\
& \text{pdf}(\Mbc)\cdot \text{pdf}(\De)\cdot \text{pdf}(\CS),\\
\end{split}
\end{equation}
where $\mathcal{F}$ represents the time-independent PDF, and pdf is the PDF used to model each variable as described below. The function $\mathcal{P}$ represents the time-dependent PDF, also described below.

Correlations among observables are considered as a source of systematic uncertainty.
    The largest linear correlation is between \Mbc and \De (10\% for \BzchInoSpace\  and 20\% for \BzchTnoSpace, respectively),
    due to the presence of photons in the final state. Linear correlations between other observables are smaller than 5\%.

The \Mbc distribution is modeled with a sum of two Gaussian functions with a common mean for signal,
a Crystal Ball function~\cite{Gaiser:Phd, Skwarnicki:1986xj, PhysRevD.25.2259} for SxF,
an ARGUS function~\cite{1990PhLB..241..278A} for continuum, and an ARGUS plus a Gaussian function for \BBbar.
In the case of \De, two Gaussian functions are used for signal and SxF, with the addition of a linear function for \BzchTnoSpace.
For continuum, we use an exponential for \BzchI and the sum of an exponential and a wide Gaussian function for \BzchTnoSpace.
For \BBbar, we use an exponential plus a Gaussian function.
For \CS, the sum of an asymmetric and a regular Gaussian function is used for signal, SxF, and \BBbar, and two Gaussian functions for continuum.

In the first step of the fit, for signal PDF we fix from simulation the
    sigma of the wider Gaussian functions for \Mbc and \De, whose fractions
    are about $2\%$ and $10\%$ of the total PDF for \Mbc and \De, respectively. For \CS, we fix the mean
    and sigma of the wider regular Gaussian function (fraction about
        $20\%$). The evaluation of the systematic uncertainties associated with fixed parameters is described in Sec.~\ref{sec:syst}.

The \Dt model for the signal and SxF components is derived from Eq.~\ref{eq:P}.
Taking into account the probability of assigning the wrong flavor, $w$, its difference between \PBz and \PaBz, $\Delta{w}$,
and the tagging efficiency asymmetry for \PBz and \PaBz, \aEpsTag, Eq.~\ref{eq:P} becomes

\begin{widetext}
    
\begin{equation} \label{eq:DtPdf}
 \begin{split}
 \mathcal{P} (&\Delta{t}, \q) =  \frac{e^{-|\Delta{t}|/\tau_{\PBz}}}{4\tau_{\PBz}}
   \Biggl\{1-\q\Delta{w}+\q\aEpsTag(1-2w)+\\
   & \left[\q(1-2w)+\aEpsTag(1-\q\Delta{w})\right] 
   \left[\SetapKs\sin(\Delta{m_d}\Delta{t})-\CetapKs\cos(\Delta{m_d}\Delta{t})\right]\Biggr\}. \\
 \end{split}
\end{equation}
\end{widetext}

The effect of detector resolution on \Dt changes Eq.~\ref{eq:DtPdf} to

\begin{equation}\label{eq:Psig}
    \mathcal{P}_{\rm exp}(\Delta{t},\q) = \int \mathcal{P}(\Delta{t'},\q) \mathcal{R}(\Delta{t}-\Delta{t'}| \sigma_{\Delta{t}}) d\Delta{t'},
\end{equation}
where $\mathcal{R}$ is the resolution function for \Dt conditional on \sDt.
The $\mathcal{R}$ function has been determined in data using $\PBz\to\PD^{(*)-}\Pgpp$ decays and it is described in details in Ref.~\cite{PhysRevD.107.L091102}.
Like the resolution function, also the flavor tagging parameters $w$, $\Delta{w}$, and \aEpsTag\ are extracted from data using flavor-specific $\PBz\to\PD^{(*)-}\Pgpp$ decays.
In simulation all these parameters from $\PBz\to\PD^{(*)-}\Pgpp$ are compatible with our expectations based on signal.

The \Dt distribution for the continuum component is modeled with three Gaussian functions and
is determined using the data sidebands described earlier.
The \Dt distribution of the \BBbar background is also modeled with the sum of three Gaussian functions and a component with an effective lifetime, that accounts for the sizable \PB lifetime, convolved with the same three Gaussian functions. Its parameters are determined from simulation.

The complete PDF used for the likelihood fit is the following:
\begin{widetext}
\begin{equation}\label{eq:pdfFull}
\begin{split}
    \text{pdf}\biggl(\Mbc, \De&, \CS ,  \Delta{t}, q_{tag}; \CetapKs, \SetapKs  \biggr) = \\
    Y_{\text{sig}}\biggl[&\bigl\{\mathcal{F}_\text{sig}(\Mbc,\De,\CS)+
    f_\text{SxF}\mathcal{F}_\text{SxF}(\Mbc,\De,\CS) \bigr\}\cdot
    \mathcal{P}_\text{exp}(\Delta{t}, q_\text{tag} ;\CetapKs,\SetapKs) \biggr] +\\ 
    Y_\text{cont}\biggl[&\mathcal{F}_\text{cont}(\Mbc,\De,\CS)\mathcal{P}_\text{cont}(\Delta{t}) +  
    f_{\BBbar}\mathcal{F}_{\BBbar}(\Mbc,\De,\CS)\mathcal{P}_{\BBbar}(\Delta{t})\biggr],\\
\end{split}
\end{equation}
\end{widetext}
where $Y_\text{sig/cont}$ is the yield of the corresponding component, $f_\text{SxF}$ is the fraction of SxF with respect to  signal, and $f_{\BBbar}$ is the fraction of \BBbar with respect to continuum.
Combined PDFs for time-independent observables are represented by $\mathcal{F}_\text{x}$ (Eq.~\ref{eq:TIPDF}) for each component x, and the time-dependent one by $\mathcal{P}_\text{x}$ (Eq.~\ref{eq:Psig}).

To validate the resolution function, we determine the \PB meson's lifetime on data,
using a simultaneous fit to both \Petaprime subchannels, separately for charged $\PBp\to\Petaprime\PKp$ and neutral $\PBz\to\Petaprime\PKs$ decays.
The results are $\tau_{\PBp}=1.63\pm0.04\ps$ and 
$\tau_{\PBz}=1.55\pm0.07\ps$, where the uncertainties are statistical.
Both are in agreement with their world averages~\cite{ParticleDataGroup:2022pth}. 

\begin{figure}[!htb]
    \centering
    \includegraphics[width=0.9\linewidth]{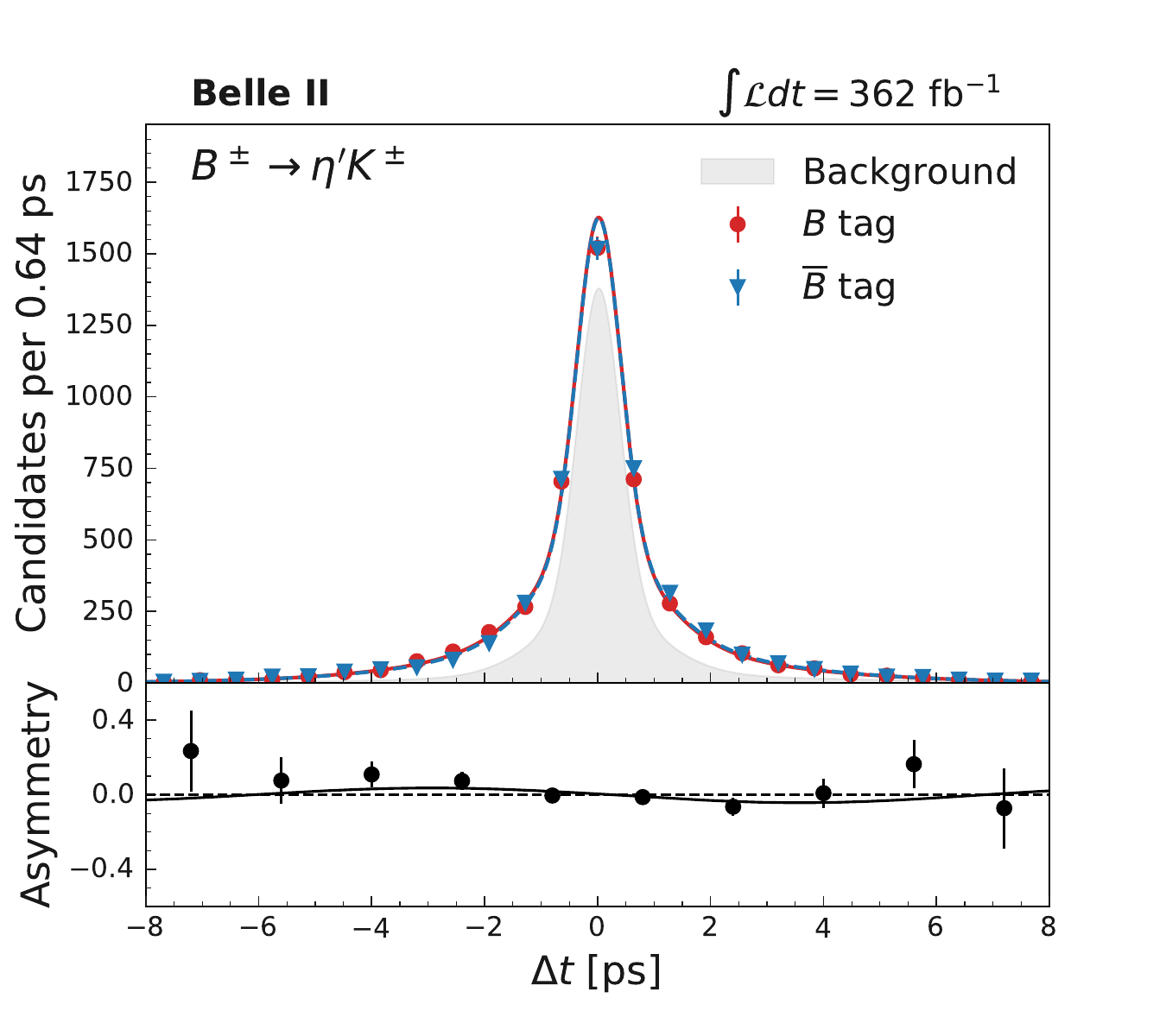}

    \caption{Distribution of \Dt for control channels {$\PBp\to\Petaprime\PKp$} separately for
    \PB and \PaB~tags, combining the two subchannels \BpchInoSpace and \BpchTnoSpace.
    The background contribution is shown as a shaded area.
    The fit projections corresponding to \PB ($\q=1$) and \PaB ($\q=-1$) 
    tags are shown as solid red and dashed blue curves, respectively.
    The bottom panel shows the asymmetry as defined in Eq.~\ref{eq:asym},
    after subtracting the background using the \sPlot technique~\cite{Pivk:2004ty}.}
    \label{fig:RawAsymBpchAll}
\end{figure}

The fit is further validated measuring the \CP asymmetry on the charged $B$ decay,  which is expected to be negligible,
with a simultaneous fit to the two \Petaprime subchannels.
The same flavor tagger used for the neutral channels is used also for the control ones.
In the signal region we find $1345\pm39$ and $1694\pm64$ signal events for \BpchInoSpace\ and \BpchTnoSpace\ subchannels
with purities (which is the ratio of signal events over the total number of events in the signal region) $77\%$ and $24\%$, respectively. 
The resulting \CP violation parameters are $\CetapKp={-0.018\pm0.044}$ and $\SetapKp=-0.083\pm0.059$, where
the uncertainties are statistical. The results are consistent with expectations of zero \CP asymmetry. 
The \Dt distributions are shown separately for \PB and \PaB tagged events in Fig.~\ref{fig:RawAsymBpchAll}, along with the asymmetry as defined in Eq.~\ref{eq:asym}.

\begin{figure}[!htb]
    \centering
    \includegraphics[width=0.382\textwidth]{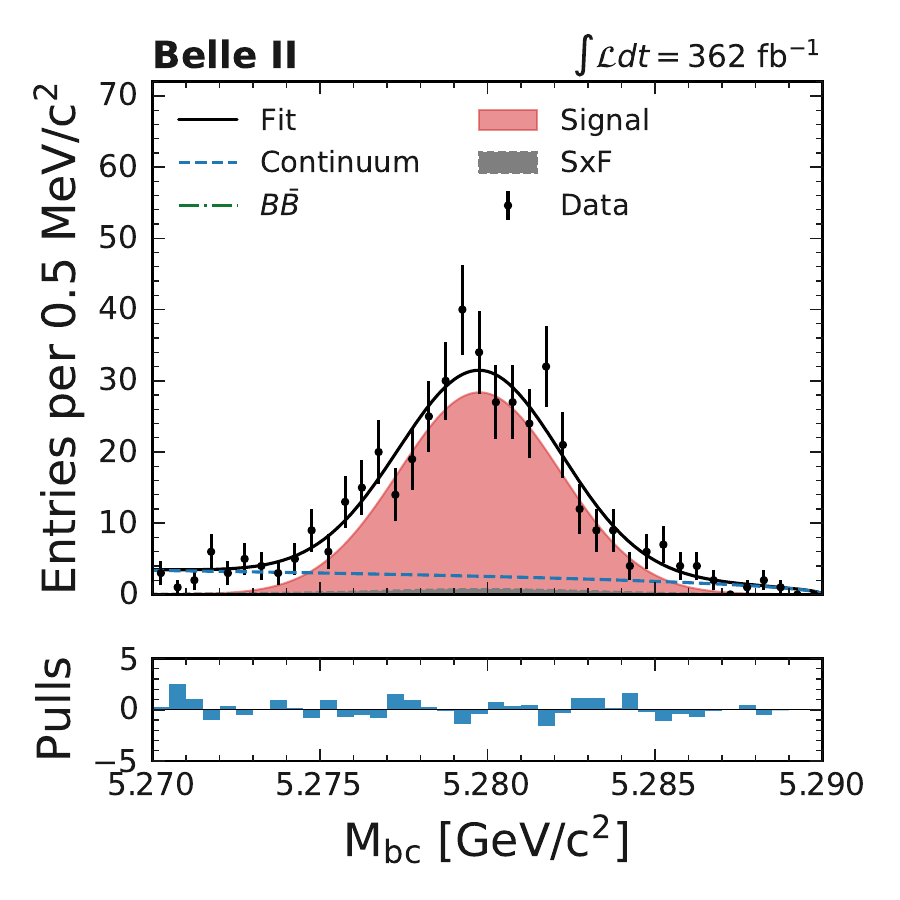}
    \includegraphics[width=0.382\textwidth]{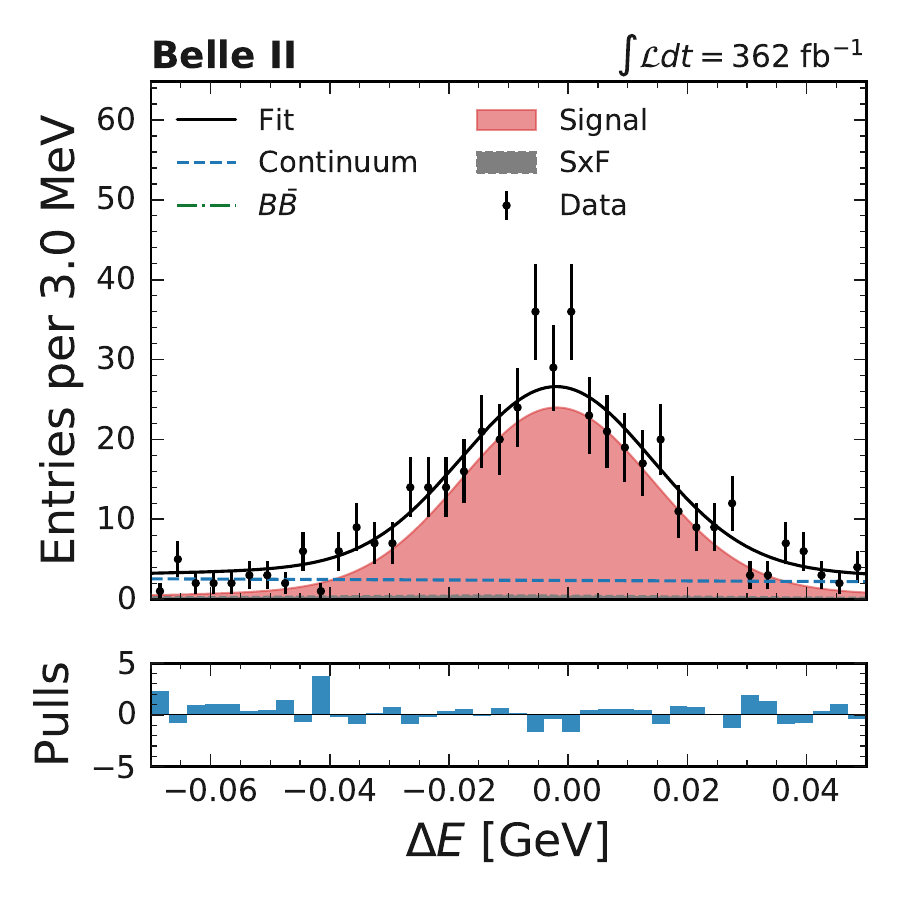}
    \includegraphics[width=0.382\textwidth]{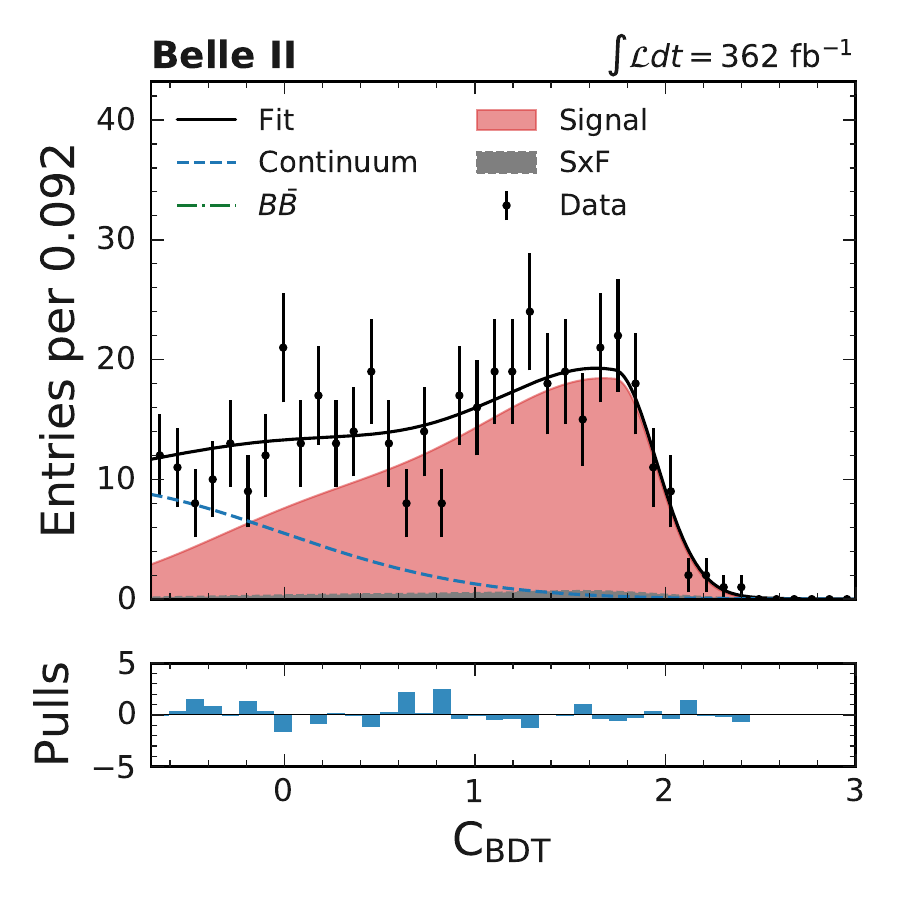}
    \caption{Distributions of \Mbc, \De, and \CS on data for \BzchInoSpace, with fit projections overlaid.
    The bottom panel shows the pull, which is the difference between data and fit, normalized to the statistical uncertainty on data.}
    \label{fig:B0ch1}
\end{figure}

\begin{figure}[!htb]
    \centering
    \includegraphics[width=0.384\textwidth]{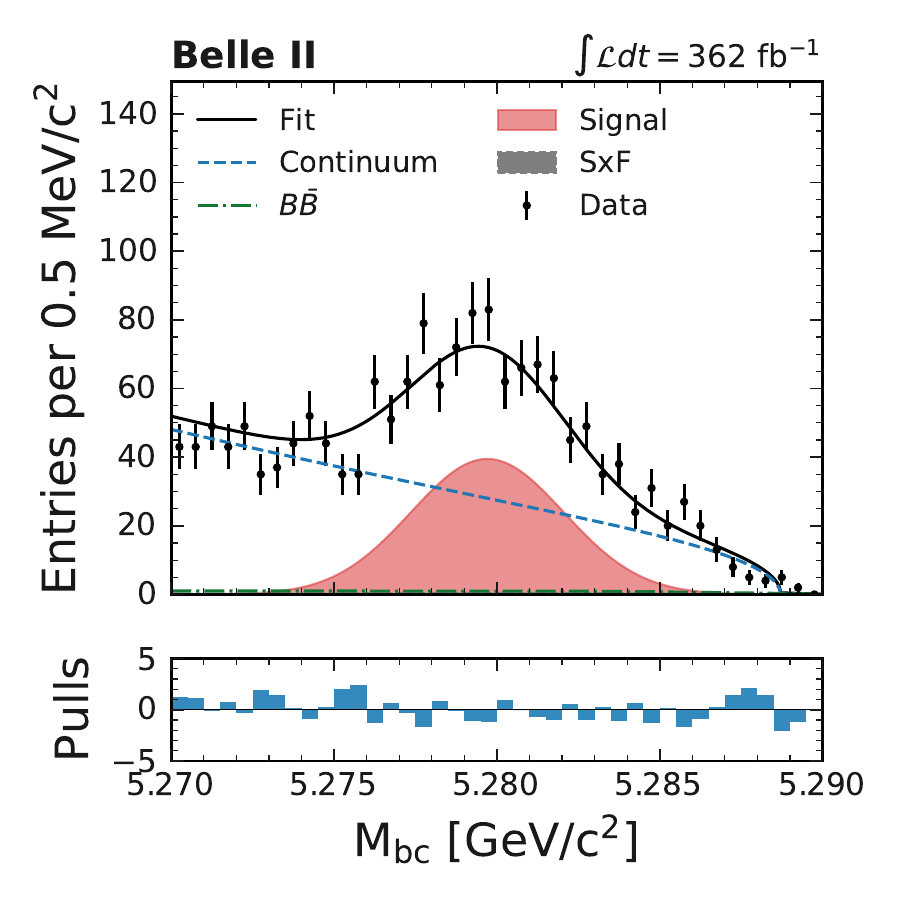}
    \includegraphics[width=0.384\textwidth]{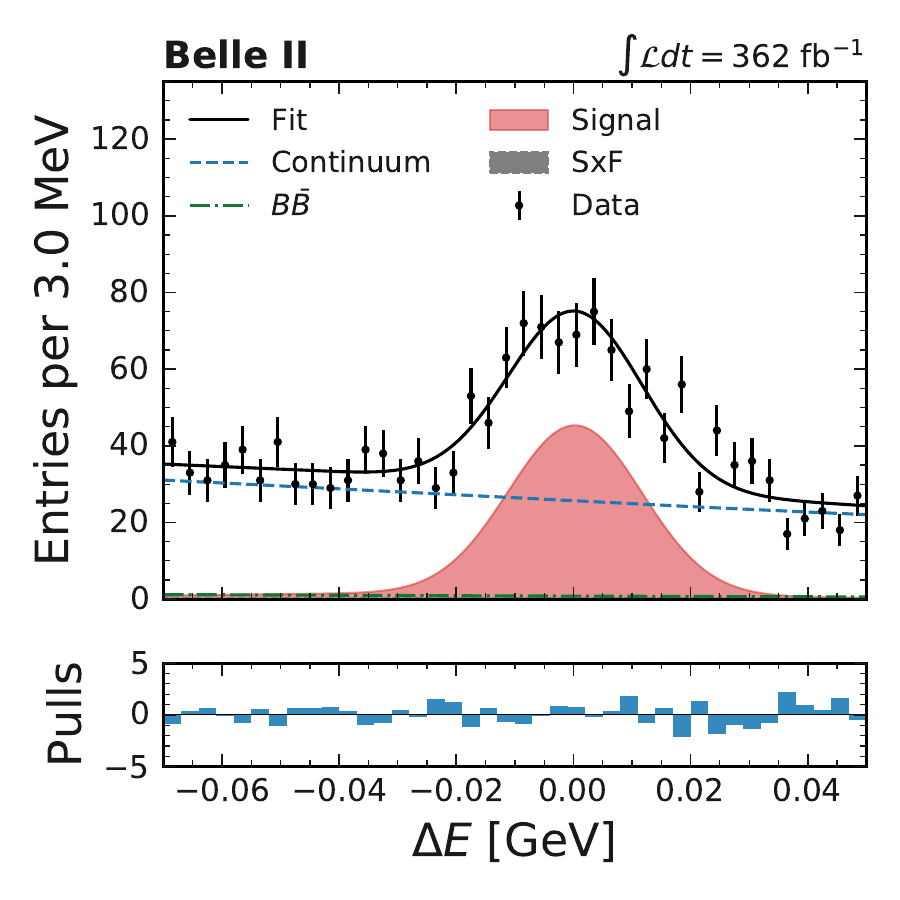}
    \includegraphics[width=0.384\textwidth]{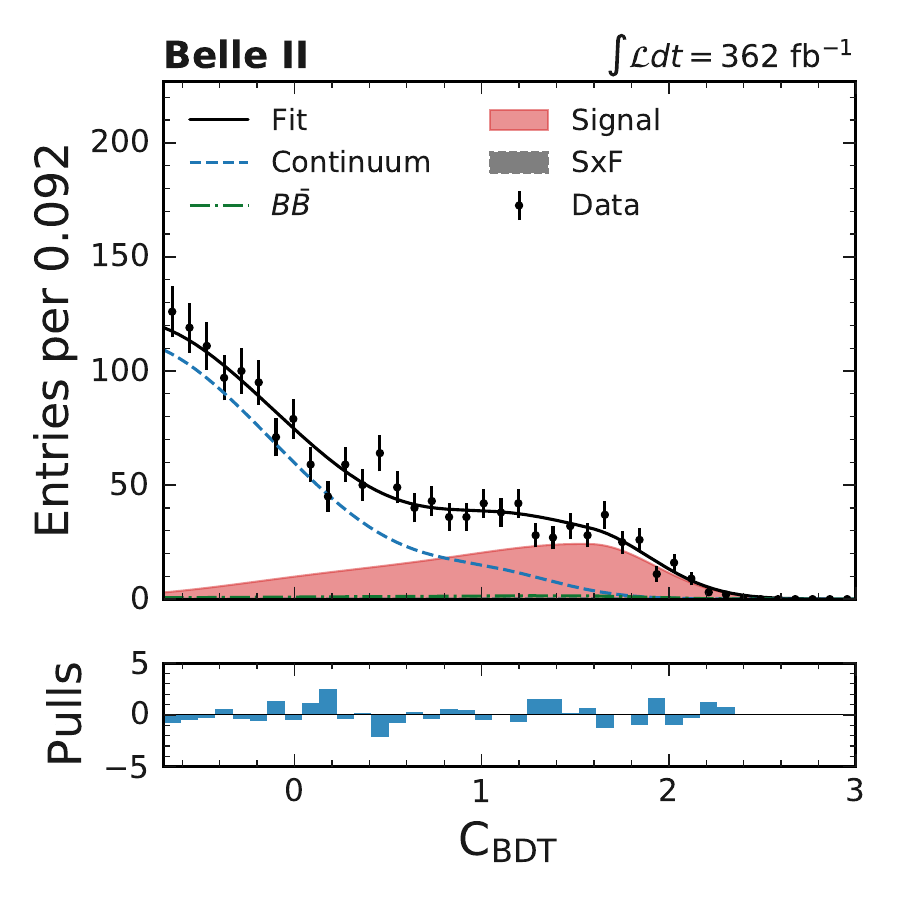}
    \caption{Distributions of \Mbc, \De, and \CS on data for \BzchTnoSpace, with fit projections overlaid.
    The bottom panel shows the pull.\\}
    \label{fig:B0ch3}
\end{figure}

\begin{figure}[!htb]
    \centering
    \includegraphics[width=0.9\linewidth]{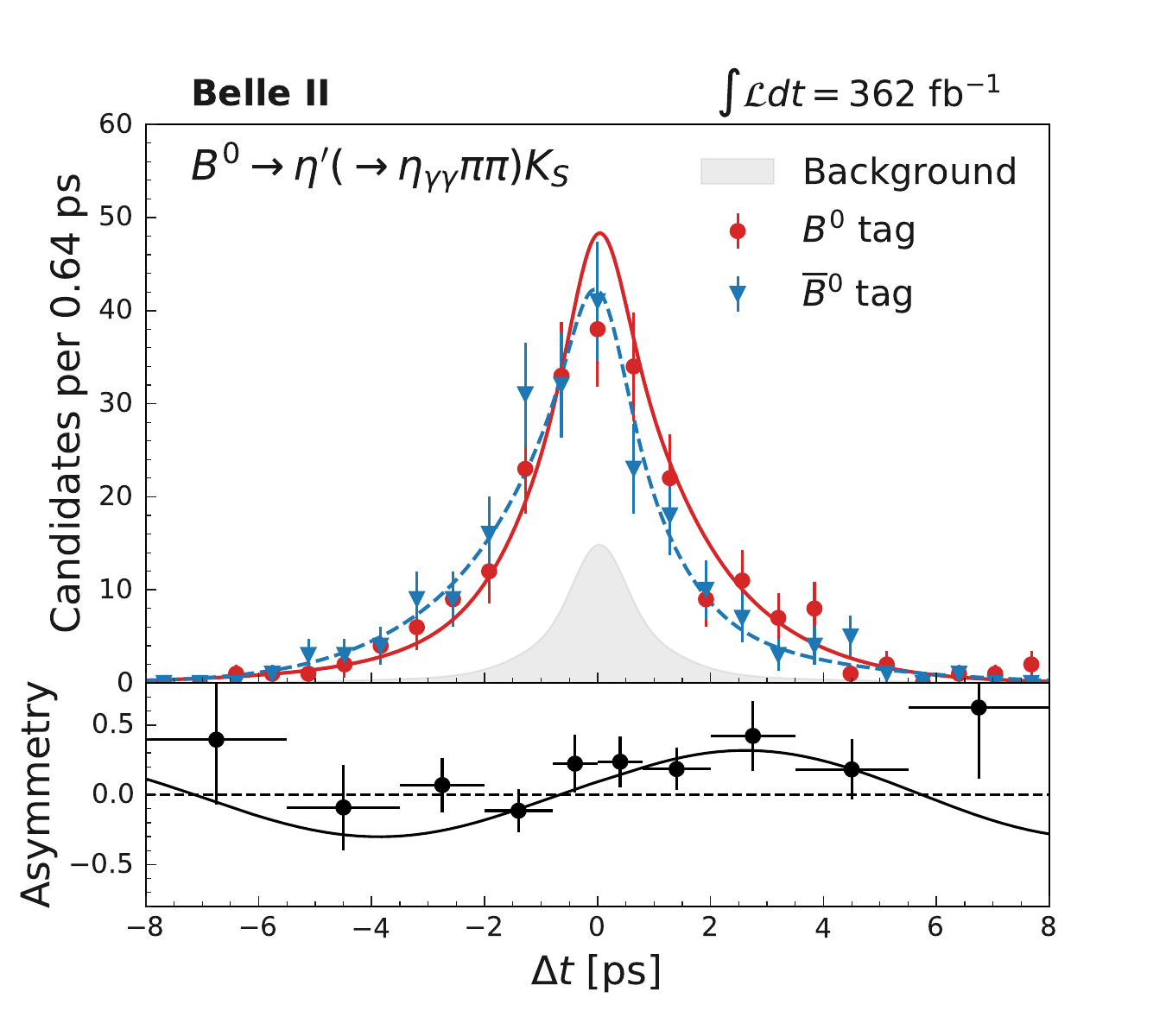}

    \caption{Distribution of \Dt for signal channels separately for \PBz and \PaBz tags, for \BzchInoSpace.
    The background contribution is shown as a shaded area.
    The fit projections  corresponding to \PBz ($\q=+1$) and \PaBz ($\q=-1$) are shown as solid and dashed curves, respectively.
    The bottom panel shows the asymmetry as defined in Eq.~\ref{eq:asym},
    after subtracting the background using the \sPlot technique~\cite{Pivk:2004ty}.
    The horizontal bars represent the widths of the variable-size bins used.
    }
    \label{fig:RawAsymB0ch1}
\end{figure}

\begin{figure}[!htb]
    \centering
    \includegraphics[width=0.9\linewidth]{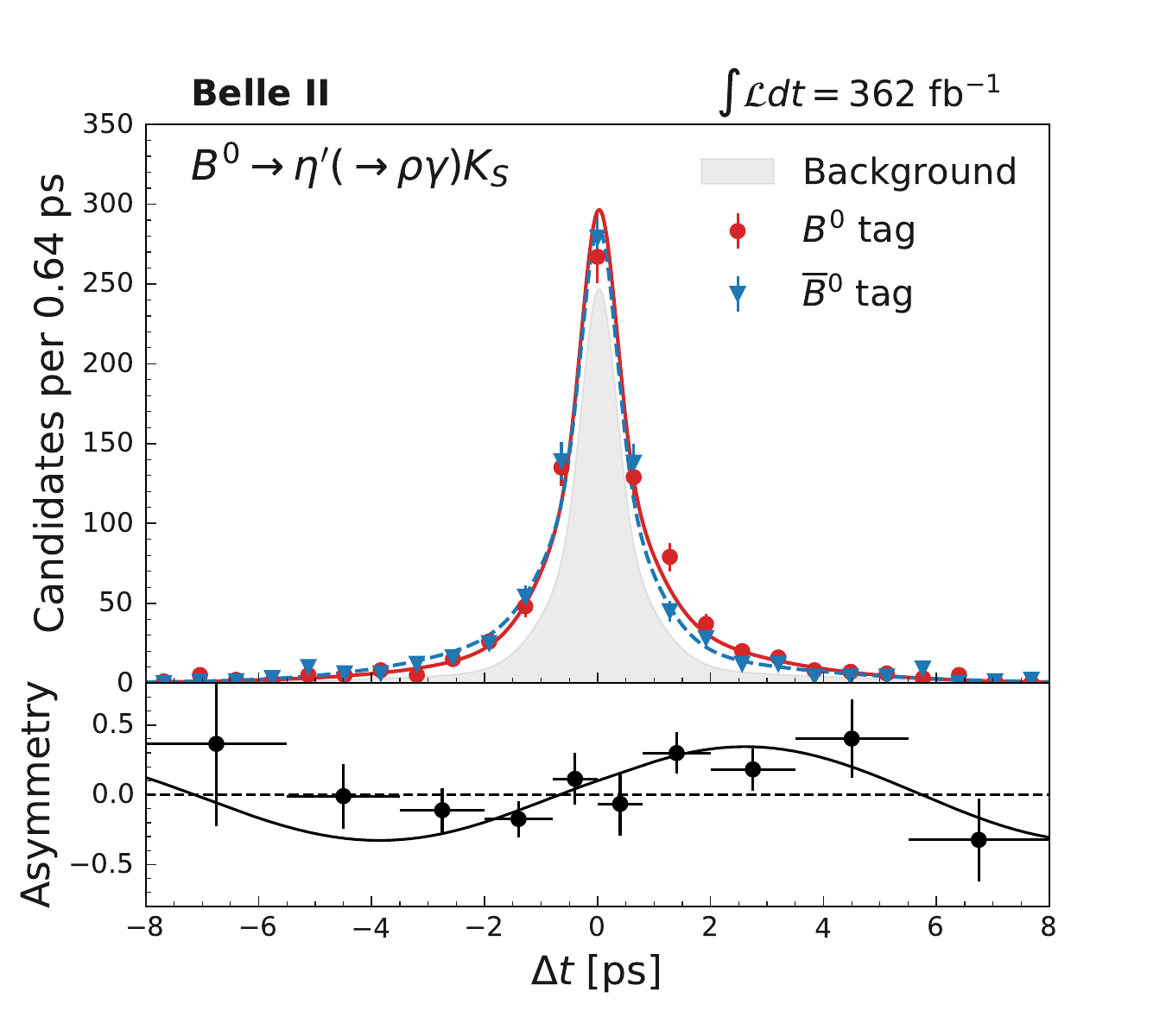}

    \caption{Distribution of \Dt for signal channels separately for \PBz ($\q=+1$)  and \PaBz ($\q=-1$) tags, for \BzchTnoSpace.
    The background contribution is shown as a shaded area.
    The fit projections corresponding to \PBz and \PaBz are shown as solid and dashed curves, respectively.
    The bottom panel shows the asymmetry as defined in Eq.~\ref{eq:asym},
    after subtracting the background using the \sPlot technique~\cite{Pivk:2004ty}.
    The horizontal bars represent the widths of the variable-size bins used.
    }
    \label{fig:RawAsymB0ch3}
\end{figure}

\begin{figure}[!htb]
    \centering
    \includegraphics[width=0.9\linewidth]{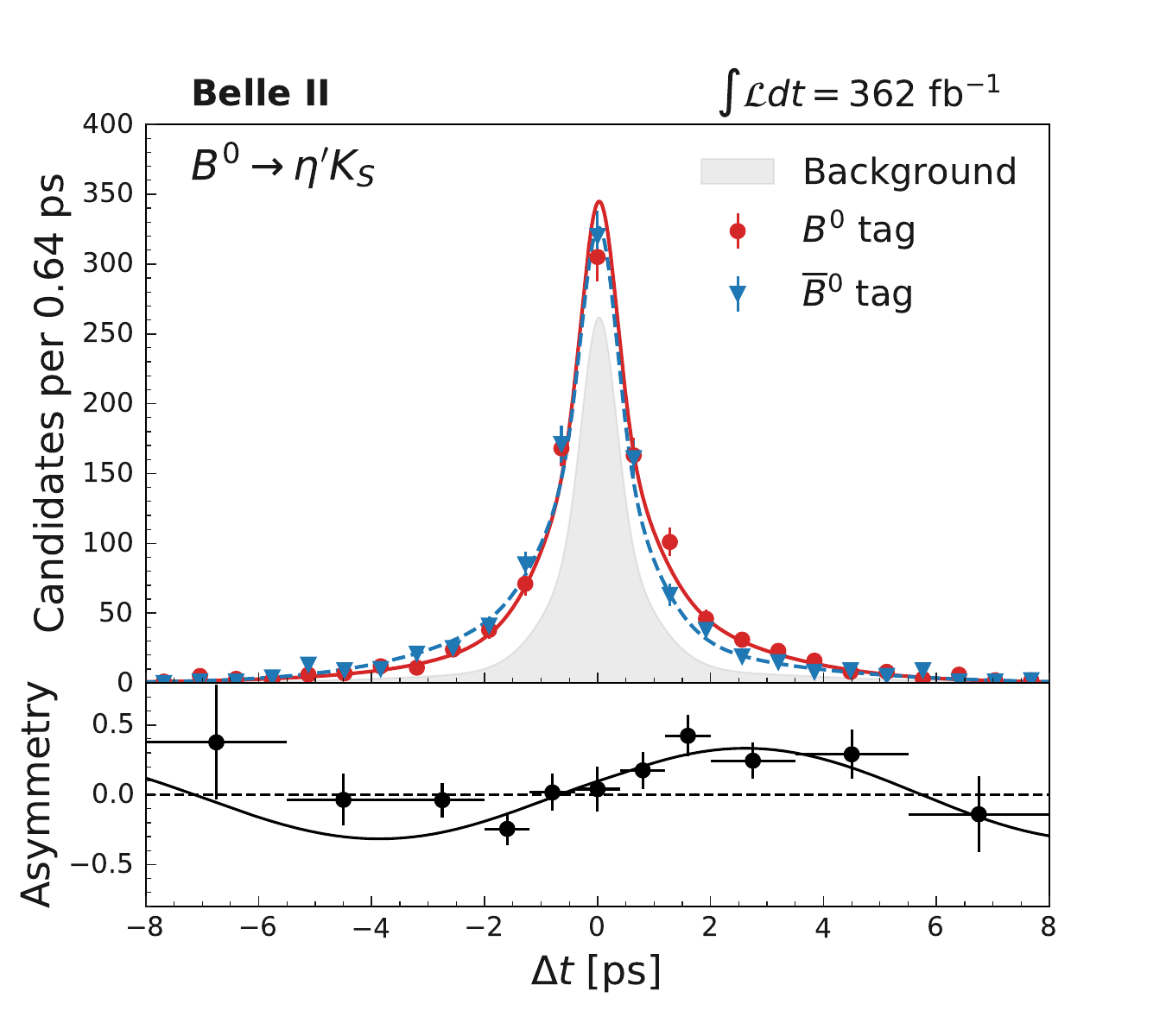}
    \caption{Distribution of \Dt for signal channels separately for \PBz ($\q=+1$) and \PaBz ($\q=-1$) tags, combining the two subchannels, \BzchInoSpace and \BzchTnoSpace.
    The background contribution is shown as a shaded area.
    The fit projections corresponding to \PBz and \PaBz are shown as solid and dashed lines, respectively.
    The bottom panel shows the asymmetry as defined in Eq.~\ref{eq:asym},
    after subtracting the background using the \sPlot technique~\cite{Pivk:2004ty}.
    The horizontal bars represent the widths of the variable-size bins used.
    }
    \label{fig:RawAsym}
\end{figure}

We then apply the fit to our signal, the neutral $B$ samples. 
The distributions of the fit observables, \Mbc, \De, and \CS in the signal region are shown in Fig.~\ref{fig:B0ch1} for the \BzchInoSpace\ subchannel, and in Fig.~\ref{fig:B0ch3} for \BzchTnoSpace\ together with the fit results of the four components. 
The \Dt distributions are shown separately for \PBz and \PaBz tagged events in Figs.~\ref{fig:RawAsymB0ch1} and~\ref{fig:RawAsymB0ch3} for 
subchannels \BzchInoSpace\ and \BzchTnoSpace, respectively, and in Fig.~\ref{fig:RawAsym} for both, along with the asymmetry as defined in Eq.~\ref{eq:asym}.


The resulting signal yield is \mbox{$358\pm20$} for \BzchInoSpace, and \mbox{$471\pm29$} for \BzchTnoSpace.  The purities for the two subchannels are 79\% and 30\%, respectively.
Results for \CP\ parameters from a simultaneous fit of the two subchannels are \mbox{$\CetapKs=-0.19\pm0.08$} and $\SetapKs=+0.67\pm0.10$, 
where the uncertainties are statistical and obtained from a scan of the likelihood ratio. The results for individual channels are summarized in Table~\ref{tab:CPres}. The correlation between \CetapKs and \SetapKs is $+3.4\%$.

\begin{figure}[!htb]
    \centering
    \includegraphics[width=0.9\linewidth]{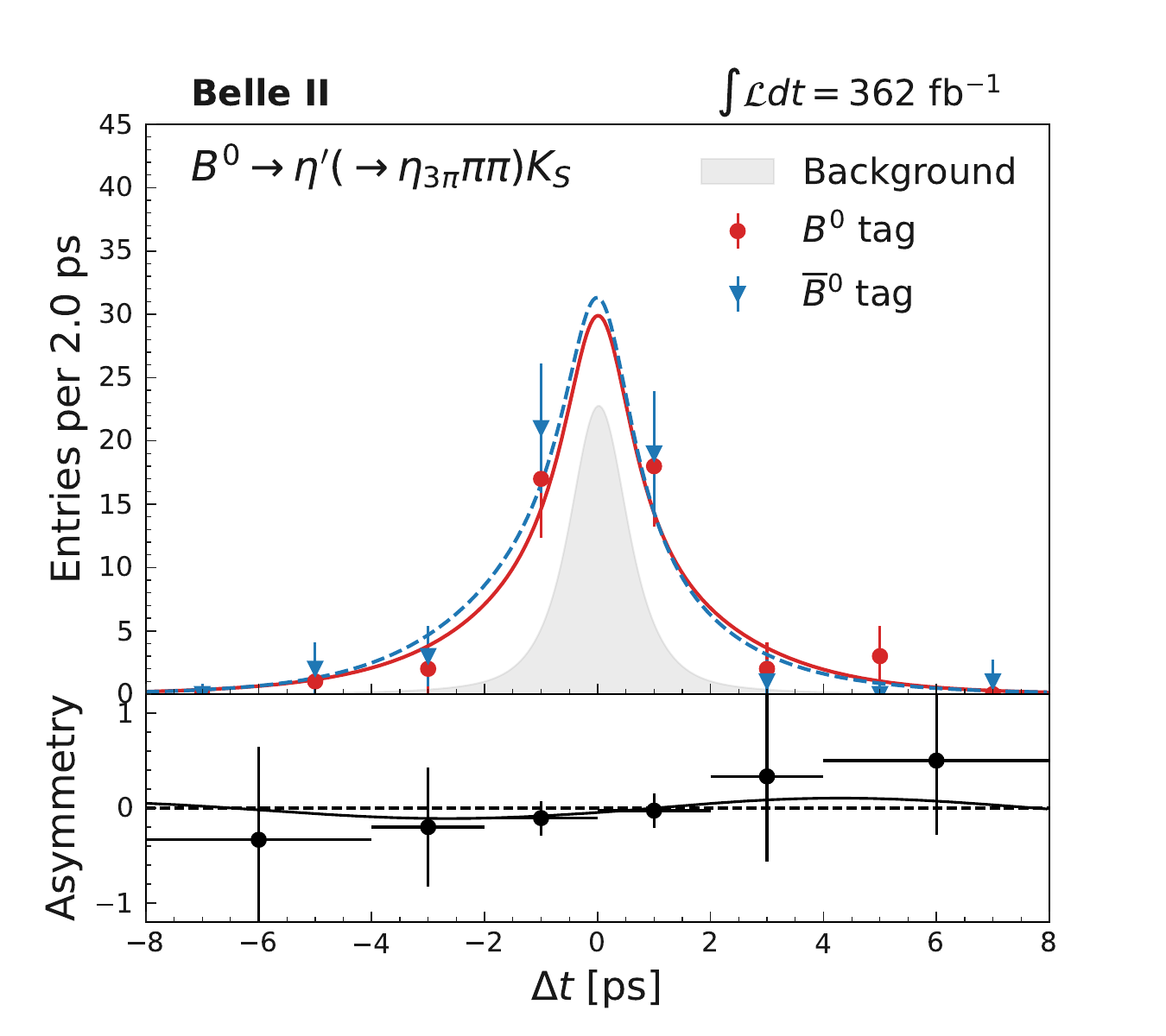}
    \caption{Distribution of \Dt for the $\PBz \to \etapr [\to \eta(\to \pip \pim \piz) \pip \pim] \KS$ channel, separately for \PBz ($\q=+1$) and \PBz ($\q=-1$) tagged events.
    The fit projections corresponding to \PBz and \PaBz are shown as solid and dashed lines, respectively.
    The background is shown as a shaded area.
    The bottom panel shows the asymmetry as defined in Eq.~\ref{eq:asym},
    using the \sPlot technique~\cite{Pivk:2004ty} to subtract the background.
    The horizontal bars represent the widths of the variable size bins used.}
    \label{fig:eta3pi_RawAsym}
\end{figure}

We also explore the subchannel \mbox{$\PBz\to\Petaprime(\to\Peta\Pgpp\Pgpm)\PKs$} \mbox{with \Peta(\to\Pgpp\Pgpm\Pgpz)}.
We reconstruct this final state combining four charged pions, selected using the same criteria for the second subchannel, and one neutral pion decaying to a pair of photons, with an invariant mass $0.120<m(\gamma\gamma)<0.145 \gevcc$.
We require $E_\gamma > 80 \mev$ for photons detected in the forward region, $30 \mev$ for the barrel region, and $60 \mev$ for the backward region, to account for the different levels of background in the three ECL regions.
The selection on $E_\gamma$ are looser than for the other subchannels to improve efficiency.
The $\eta$ candidate is required to have an invariant mass in the range $0.52 < m(\pip \pim \piz) < 0.57\gevcc$ while the mass difference $\Delta{m}$ between $\etapr$ and $\eta$ candidates must satisfy $0.40 < \Delta m < 0.42 \gevcc$.
The other selection criteria for the \KS and \Bz candidates are the same as for the other subchannels.
The signal yield is $55 \pm 8$, and the purity is 55\%.
We perform the vertex fit and determine the resolution model, whose functional form is similar to that used in Ref.~\cite{TAJIMA2004370}.
The measured \CP asymmetries, $C_{\etapr(3\pi) \KS} = 0.11 ^{+0.32}_{-0.31}$ and $S_{\etapr(3\pi) \KS}=0.25 ^{+0.47}_{-0.53}$ (the uncertainties are statistical) are consistent with those determined in the other two subchannels, but significantly less precise.
Figure~\ref{fig:eta3pi_RawAsym} shows the results of the \CP asymmetry fit on the $\PBz \to \etapr [\to \eta(\to \pip \pim \piz) \pip \pim] \KS$ subchannel. The results of this subchannel are not used for final results since their statistical significance is negligible.

\begin{table}[!htb]
    \caption{Summary of results on \CetapKs and \SetapKs for the three subchannels, identified by the \Petaprime decay. The last subchannel is not included in the simultaneous fit. The uncertainties are statistical only.}
        \label{tab:CPres}
        \centering
        \begin{tabular}{l@{\hskip 3mm}r@{\hskip 3mm}c@{\hskip 3mm}@{\hskip 3mm}c@{\hskip 3mm} }
            \hline
            \hline
            {Channel} &  Signal yield & {\CetapKs} & {\SetapKs} \\
            \hline
            \hline
            $\Petaprime\to\Peta_{\Pphoton\Pphoton}\Pgpp\Pgpm$  & $358\pm20$ & $-0.10 \pm 0.13$ & $0.69\pm0.14$ \\
            $\Petaprime\to\Prho\Pphoton$  & $471\pm29$ & $-0.24 \pm 0.10$ & $0.65\pm0.13$ \\
            $\Petaprime\to\Peta_{3\Pgp}\Pgpp\Pgpm$  & $55\pm8~\,$   & $~\,\,0.11 \pm 0.32$ & $0.25\pm0.50$\\
            \hline
            Sim. fit & $829\pm35$ & $-0.19\pm0.08$ & $0.67\pm0.10$ \\
            \hline
            \hline
  \end{tabular}
\end{table}

\section{Systematic uncertainties}\label{sec:syst}
We consider several sources of possible systematic uncertainties, which are listed in Table~\ref{tab:CPsyst}.

\begin{table}[hbt]
    \caption{Summary of systematic uncertainties for \CetapKs and \SetapKs.}
        \label{tab:CPsyst}
        \centering
        \begin{tabular}{l@{\hskip 3mm}r@{\hskip 3mm}@{\hskip 3mm}r@{\hskip 3mm} }
        \hline
        \hline
            {Source} &  {\CetapKs} & {\SetapKs} \\
            \hline
            Signal and continuum yields                & $<0.001$ & 0.002 \\
            SxF and \BBbar yields                      & $<0.001$ & 0.006 \\
            \CS mismodeling                            & 0.004    & 0.010 \\
            Signal and background modeling            & 0.011    & 0.011 \\
            Observable correlations                     & 0.008    & 0.001 \\
            $\Delta{t}$ resolution fixed parameters    & 0.005    & 0.009 \\
            $\Delta{t}$ resolution model               & 0.004    & 0.019 \\
            Flavor tagging                             & 0.007    & 0.004 \\
            $\tau_{\PBz}$ and $\Delta{m_d}$               & $<0.001$ & 0.002 \\
            Fit bias                                   & 0.003    & 0.002 \\
            Tracker misalignment                       & 0.004    & 0.006 \\
            Momentum scale                             & 0.001    & 0.001 \\
            Beam spot                                  & 0.002    & 0.002 \\
            $B$-meson motion in the $\PupsilonFourS$ frame & $<0.001$ & 0.017 \\
            Tag-side interference                      & 0.027    & $<0.001$ \\
            \BBbar background  asymmetry               & 0.008    & 0.006 \\
            Candidate selection                        & 0.007    & 0.009 \\

            \hline
            Total                                      & 0.034   & 0.034 \\
            \hline
            \hline
  \end{tabular}
\end{table}

To determine most of the systematic uncertainties, we use ensembles of simulated samples.
For signal and SxF events we use sampling with replacement~\cite{10.1214/aos/1176344552} from the available simulated samples,
while continuum and \BBbar backgrounds are sampled from the PDF used for modeling, due to the limited size of simulated samples. 

To evaluate the impact of the uncertainties of signal and continuum yields in the second
fit step, we vary them individually by assuming alternate values, corresponding to
$\pm1$-standard-deviation fluctuations of the first fit results.
We then consider the difference between the values of \CetapKs and \SetapKs from the nominal results as a systematic uncertainty.

The SxF and \BBbar yields are fixed to their values obtained in simulation. To estimate the systematic uncertainties, 
we let the fit determine the \BBbar yield for all $r$ bins, one at a time to ensure the fit convergence.
The average ratio between the fit result and expectation is $0.9\pm0.2$,
and the average uncertainty of \BBbar yield in each bin is about $50\%$.
We vary the \BBbar yield by this uncertainty and take the variations in the \CP asymmetry parameters
as a systematic uncertainty.
We also vary the SxF yield by the same uncertainty (50\%), and evaluate the variations on \CP asymmetry parameters.

The impact of the choice of training sample for \CS is evaluated by comparing the results obtained using the data sidebands
for its training with those in which the training is performed on continuum simulated events.

Similarly to what is done for yields, the parameters of PDF shapes used 
to model the various components are varied within their
statistical uncertainties if they are determined in the first step of the fit.
This is the case for the signal and continuum components.
The variations of \CetapKs and \SetapKs due to alternative values of each parameter are summed linearly,
to account for possible correlation, and used as a systematic uncertainty.
The parameters fixed from simulation are allowed to vary, individually,
    in the yield fit, and then the \CP asymmetry fit is performed using the
    varied yields as input.
We take the difference of \CetapKs and \SetapKs from the nominal results
as a systematic uncertainty, summing the differences in quadrature.

The impact of the correlation between the fit observables, mostly between \Mbc and \De,
is estimated using ensembles of signal events simulated with correlation,
sampled with replacement from the available simulated samples,
and without correlation, by sampling the PDFs, and comparing the two sets of results.

The systematic uncertainty due to the \Dt resolution model is estimated by repeating the fit under
alternative assumptions for values of
the resolution parameters within the uncertainties obtained from the fit on the $\PB\to\PD^{(*)-}\Pgpp$ control sample.
We also compare the result of a fit using the resolution parameters obtained by fitting the simulated \mbox{$\PB\to\Petaprime\PKs$} signal events with the nominal fit, and take the difference on \CetapKs and \SetapKs as a systematic uncertainty.

The flavor-tagging parameters are varied within their statistical uncertainties.
The same is done for the physics parameters $\tau_{\PBz}$ and $\Delta{m_d}$, using the uncertainties on their world-average values.
Small fit biases found on \CetapKs and \SetapKs by fits to large simulated samples are also included.

The impact of tracker misalignment is estimated with dedicated simulation samples with four different misalignment scenarios.
The effects on charged-particle momentum scale due to imperfect modeling of the magnetic field,
and the uncertainty on the beam spot determination, were studied in a previous analysis 
that used similar reconstruction strategies~\cite{JspiK}. 

We estimate the effect of neglecting the small motion of the \PB meson in the \PupsilonFourS rest
frame by calculating \Dt from simulated signal events and comparing the true \Dt with that based on true $\Delta{z}$.

The \Dt model in Eq.~\ref{eq:asym} assumes that the \Btag decays into a flavor-specific mode.
The impact of tag-side interference, namely the presence of Cabibbo-suppressed $b\to{u\bar{c}s}$ decays 
in the \Btag with different weak phase~\cite{PhysRevD.68.034010}, introduces a systematic which has been evaluated
as described in~\cite{PhysRevLett.108.171802}.
We conservatively assume that all events are tagged by hadronic \PB decays, where the effect is largest.
We use the difference with respect to the nominal asymmetry as a systematic uncertainty.

The \BBbar background is small and dominated by $b\to{c}$ decays. 
The possible presence of \CP violation in the \BBbar background is estimated conservatively by assuming that 
all this background has a \CP asymmetry with $(\Ccp,\Scp)=(\pm0.2,0)$ or $(0,\pm0.2)$ and taking the
largest variation on \CetapKs and \SetapKs as a systematic uncertainty.

Finally, the impact of the candidate selection is evaluated by repeating the
full analysis with all multiple candidates and comparing with the results obtained using only the one with best vertex-$\chi^2$.

\section{Summary}
A measurement of \CP asymmetries in $\PB\to\Petaprime\PKs$ decays
is conducted using $e^+e^-$ collision data collected in 2019–-2022 by the Belle II experiment at the SuperKEKB collider.
We find $829\pm 35$ signal decays in a sample of $\num{387\pm6 e6}$ \BBbar events and measure the \CP asymmetries to be

\begin{equation}
\begin{split}
    \CetapKs & = -0.19 \pm 0.08 \pm 0.03, \\ 
    \SetapKs & = +0.67 \pm 0.10 \pm 0.03, \\ 
\end{split}
\end{equation}
where the first uncertainties are statistical and the second systematic. 
This measurement is based on the subchannels \BzchInoSpace and \BzchTnoSpace.

This is the first measurement of \CP violation in this channel at Belle II.
The results are in agreement with the current world averages,
and have sensitivities close to those of Belle~\cite{Belle:2014atq} and BaBar~\cite{BaBar:2008ucf}, despite the smaller data size.

\section{Acknowledgements}
This work, based on data collected using the Belle II detector, which was built and commissioned prior to March 2019, was supported by
Higher Education and Science Committee of the Republic of Armenia Grant No.~23LCG-1C011;
Australian Research Council and Research Grants
No.~DP200101792, 
No.~DP210101900, 
No.~DP210102831, 
No.~DE220100462, 
No.~LE210100098, 
and
No.~LE230100085; 
Austrian Federal Ministry of Education, Science and Research,
Austrian Science Fund
No.~P~31361-N36
and
No.~J4625-N,
and
Horizon 2020 ERC Starting Grant No.~947006 ``InterLeptons'';
Natural Sciences and Engineering Research Council of Canada, Compute Canada and CANARIE;
National Key R\&D Program of China under Contract No.~2022YFA1601903,
National Natural Science Foundation of China and Research Grants
No.~11575017,
No.~11761141009,
No.~11705209,
No.~11975076,
No.~12135005,
No.~12150004,
No.~12161141008,
and
No.~12175041,
and Shandong Provincial Natural Science Foundation Project~ZR2022JQ02;
the Czech Science Foundation Grant No.~22-18469S;
European Research Council, Seventh Framework PIEF-GA-2013-622527,
Horizon 2020 ERC-Advanced Grants No.~267104 and No.~884719,
Horizon 2020 ERC-Consolidator Grant No.~819127,
Horizon 2020 Marie Sklodowska-Curie Grant Agreement No.~700525 ``NIOBE''
and
No.~101026516,
and
Horizon 2020 Marie Sklodowska-Curie RISE project JENNIFER2 Grant Agreement No.~822070 (European grants);
L'Institut National de Physique Nucl\'{e}aire et de Physique des Particules (IN2P3) du CNRS
and
L'Agence Nationale de la Recherche (ANR) under grant ANR-21-CE31-0009 (France);
BMBF, DFG, HGF, MPG, and AvH Foundation (Germany);
Department of Atomic Energy under Project Identification No.~RTI 4002,
Department of Science and Technology,
and
UPES SEED funding programs
No.~UPES/R\&D-SEED-INFRA/17052023/01 and
No.~UPES/R\&D-SOE/20062022/06 (India);
Israel Science Foundation Grant No.~2476/17,
U.S.-Israel Binational Science Foundation Grant No.~2016113, and
Israel Ministry of Science Grant No.~3-16543;
Istituto Nazionale di Fisica Nucleare and the Research Grants BELLE2;
Japan Society for the Promotion of Science, Grant-in-Aid for Scientific Research Grants
No.~16H03968,
No.~16H03993,
No.~16H06492,
No.~16K05323,
No.~17H01133,
No.~17H05405,
No.~18K03621,
No.~18H03710,
No.~18H05226,
No.~19H00682, 
No.~20H05850,
No.~20H05858,
No.~22H00144,
No.~22K14056,
No.~22K21347,
No.~23H05433,
No.~26220706,
and
No.~26400255,
the National Institute of Informatics, and Science Information NETwork 5 (SINET5), 
and
the Ministry of Education, Culture, Sports, Science, and Technology (MEXT) of Japan;  
National Research Foundation (NRF) of Korea Grants
No.~2016R1\-D1A1B\-02012900,
No.~2018R1\-A2B\-3003643,
No.~2018R1\-A6A1A\-06024970,
No.~2019R1\-I1A3A\-01058933,
No.~2021R1\-A6A1A\-03043957,
No.~2021R1\-F1A\-1060423,
No.~2021R1\-F1A\-1064008,
No.~2022R1\-A2C\-1003993,
and
No.~RS-2022-00197659,
Radiation Science Research Institute,
Foreign Large-Size Research Facility Application Supporting project,
the Global Science Experimental Data Hub Center of the Korea Institute of Science and Technology Information
and
KREONET/GLORIAD;
Universiti Malaya RU grant, Akademi Sains Malaysia, and Ministry of Education Malaysia;
Frontiers of Science Program Contracts
No.~FOINS-296,
No.~CB-221329,
No.~CB-236394,
No.~CB-254409,
and
No.~CB-180023, and SEP-CINVESTAV Research Grant No.~237 (Mexico);
the Polish Ministry of Science and Higher Education and the National Science Center;
the Ministry of Science and Higher Education of the Russian Federation
and
the HSE University Basic Research Program, Moscow;
University of Tabuk Research Grants
No.~S-0256-1438 and No.~S-0280-1439 (Saudi Arabia);
Slovenian Research Agency and Research Grants
No.~J1-9124
and
No.~P1-0135;
Agencia Estatal de Investigacion, Spain
Grant No.~RYC2020-029875-I
and
Generalitat Valenciana, Spain
Grant No.~CIDEGENT/2018/020;
National Science and Technology Council,
and
Ministry of Education (Taiwan);
Thailand Center of Excellence in Physics;
TUBITAK ULAKBIM (Turkey);
National Research Foundation of Ukraine, Project No.~2020.02/0257,
and
Ministry of Education and Science of Ukraine;
the U.S. National Science Foundation and Research Grants
No.~PHY-1913789 
and
No.~PHY-2111604, 
and the U.S. Department of Energy and Research Awards
No.~DE-AC06-76RLO1830, 
No.~DE-SC0007983, 
No.~DE-SC0009824, 
No.~DE-SC0009973, 
No.~DE-SC0010007, 
No.~DE-SC0010073, 
No.~DE-SC0010118, 
No.~DE-SC0010504, 
No.~DE-SC0011784, 
No.~DE-SC0012704, 
No.~DE-SC0019230, 
No.~DE-SC0021274, 
No.~DE-SC0021616, 
No.~DE-SC0022350, 
No.~DE-SC0023470; 
and
the Vietnam Academy of Science and Technology (VAST) under Grants
No.~NVCC.05.12/22-23
and
No.~DL0000.02/24-25.

These acknowledgements are not to be interpreted as an endorsement of any statement made
by any of our institutes, funding agencies, governments, or their representatives.

We thank the SuperKEKB team for delivering high-luminosity collisions;
the KEK cryogenics group for the efficient operation of the detector solenoid magnet;
the KEK computer group and the NII for on-site computing support and SINET6 network support;
and the raw-data centers at BNL, DESY, GridKa, IN2P3, INFN, and the University of Victoria for off-site computing support.

\bibliographystyle{aipnum4-1}
\bibliography{references}

\end{document}